# Detection of Compact Ultraviolet Nuclear Emission in LINER Galaxies[1]

Dan Maoz[2,3], Alexei V. Filippenko[4,7], Luis C. Ho[4],
Hans-Walter Rix[3,8], John N. Bahcall[3],
Donald P. Schneider[3,6], and F. Duccio Macchetto[5]

[1] Based on observations with the *Hubble Space Telecope* which is operated
by AURA under NASA contract NAS 5-26555
[2] School of Physics & Astronomy and Wise Observatory,
Tel-Aviv University, Tel-Aviv 69978, ISRAEL
[3] Institute for Advanced Study, Princeton, NJ 08540
[4] Department of Astronomy, University of California, Berkeley, CA 94720
[5] Space Telescope Science Institute, 3700 San Martin Dr., Baltimore, MD 21218
[6] Department of Astronomy and Astrophysics,
The Pennsylvania State University, University Park, PA 16802
[7] Presidential Young Investigator
[8] Hubble Fellow

# Abstract

Low-ionization nuclear emission-line regions (LINERs), which exist in a large fraction of galaxies, may be the least luminous manifestation of quasar activity. As such, they may allow the study of the AGN phenomenon in the nearest galaxies. The nature of LINERs has, however, remained controversial because an AGN-like nonstellar continuum source has not been directly observed in them. We report the detection of bright ($\gtrsim 2 \times 10^{-16}$ erg s$^{-1}$ cm$^{-2}$ Å$^{-1}$), unresolved (FWHM $\lesssim 0.1''$) point sources of UV ($\sim 2300$ Å) emission in the nuclei of nine nearby galaxies. The galaxies were imaged using the Faint Object Camera on the *Hubble Space Telescope* (*HST*), and seven of them are from a complete sample of 110 nearby galaxies that was observed with *HST*. Ground-based optical spectroscopy reveals that five of the nuclei are LINERs, three are starburst nuclei, and one is a Seyfert nucleus. The observed UV flux in each of the five LINERs implies an ionizing flux that is sufficient to account for the observed emission lines through photoionization. The detection of a strong UV continuum in the LINERs argues against shock excitation as the source of the observed emission lines, and supports the idea that photoionization excites the lines in at least some objects of this class.

We have analyzed ground-based spectra for most of the Northern-hemisphere galaxies in the *HST* sample, and find that 26 of them are LINERs, among which only the above five LINERs have a detected nuclear UV source. There are no obvious differences in the optical line intensity ratios between the UV-bright and UV-dark LINERs. If all LINERs are photoionized, then the continuum source is unobscured along our line of sight in $5/26 \approx 20\%$ of LINERs. Alternatively, it can be argued that spectrally-similar LINERs are produced by various excitation mechanisms, and that photoionization is responsible in only about 20% of the cases. The high angular resolution allows us to set upper limits, typically several parsecs, on the physical size of the compact star-cluster or AGN-type continuum source that is emitting the UV light in these objects.

*Subject headings: galaxies: active – galaxies: nuclei – ultraviolet: galaxies*

# 1   Introduction

The active galactic nucleus (AGN) phenomenon, which finds its most dramatic expression in quasars, extends to low luminosities (see Filippenko 1989 for a review; we will use the term "AGN" in this paper to mean the type of emission process occuring in quasars which, based on energetics, cannot be explained by stellar processes). The detection and quantification of the low-luminosity end of the AGN phenomenon is important for understanding the nature of AGNs and their evolution. The least luminous "classical" Seyfert 1 nucleus is that of NGC 4051, with $M_B \approx -16$ mag. (For comparison, Markarian Seyfert 1 nuclei typically have $-23 \leq M_B \leq -18$ mag, e.g. Osterbrock & Martel 1993.) A substantial number of galaxies exhibiting weak broad H$\alpha$ emission qualitatively similar to that in quasars has also been found (Stauffer 1982; Keel 1983; Filippenko & Sargent 1985). The broad H$\alpha$ emission-line luminosity is sometimes only $\sim 1/100$th that in NGC 4051, suggesting the presence of a low-luminosity active nucleus. Also possibly belonging to the AGN category are a class of objects with narrow emission lines named low-ionization nuclear emission-line regions (LINERs; Heckman 1980). The emission-line spectrum of LINERs, which exist in a large fraction of all luminous galaxies, is probably best explained by photoinization by an AGN-like nonstellar continuum (Ho, Filippenko, & Sargent 1993, and references therein). LINERs could therefore be the least luminous manifestation of the AGN phenomenon. Due to the indirectness of this inference, there has been continued debate whether LINERs are in fact AGNs. In particular, it has been suggested (e.g., Terlevich & Melnick 1985; Filippenko & Terlevich 1992; Shields 1992) that photoionization by a cluster of hot stars can reproduce the observations in some or all LINERs, or that shock excitation (e.g., Koski & Osterbrock 1976; Fosbury et al. 1978; Heckman 1980) is responsible for the LINER spectrum, and that hence these objects are not true AGNs.

One of the clearest signatures of an AGN is a compact source of nonstellar continuum emission. Direct detection of this component in LINERs could rule out the shock excitation scenario. In low-luminosity AGNs, however, any nonstellar continuum is severely diluted at optical wavelengths by light from stars in the nucleus, making the nonstellar light difficult to detect. Even in the $U$ band, where the seeing is rarely better than $1''$, starlight contamination within the seeing disk is severe. While a nonstellar continuum from low-luminosity AGNs is unobservable at optical bands, it can easily be detected in the ultraviolet (UV), where the background from the late stellar population in the bulge is low (e.g., Bohlin et al. 1985). The contrast between an AGN with an $f_\nu \propto \nu^{-1}$ spectrum and the background from an Sab galaxy (Coleman, Wu, & Weedman 1980) increases by a factor of about 15 when going from the $U$-band to 2200 Å, and by about 150 when going from the $V$-band to 2200 Å. For example, the AGN with the lowest luminosity known ($M_B = -10$ mag) is in the nearby late-type spiral galaxy NGC 4395 (Filippenko & Sargent 1989; Filippenko, Ho, & Sargent 1993) and was discovered optically only because this galaxy has practically no bulge. Assuming this AGN has a $\nu^{-1}$ continuum, at 2200 Å it would appear as an 18.5 AB magnitude point source (AB$= -48.6 - 2.5 \log f_\nu$[erg s$^{-1}$ cm$^{-2}$ Hz$^{-1}$]; Oke & Gunn 1983), easily discerned above a more typical bulge background of 19.5 AB magnitudes/arcsec$^2$.

We have imaged in the UV with the *Hubble Space Telescope* (*HST*) 110 galactic nuclei chosen randomly from a complete sample of 213 large, nearby galaxies. The principal purpose of this program is to detect low-luminosity AGNs in the UV, and to study them at small physical scales with *HST*'s high angular resolution. In this paper, we present data for the nine galaxies having the brightest UV sources in the nuclear region which we have imaged, the central $22'' \times 22''$. A point-spread-function (PSF) analysis shows that in eight of the nine cases the UV source is unresolved by *HST*, corresponding to upper limits on the physical scales of order a few parsecs. From ground-based spectroscopy, we find that five of these nine galaxies have LINER spectra. The observed strength of the UV continuum implies an ionizing continuum that is sufficient to excite the observed emission-line flux through photoionization. The data therefore demonstrate the existence of a bright UV continuum in some LINERs, and argue against the shock-excitation mechanism in these particular cases. The fact that bright UV sources are detected in only five out of 26 galaxies in our sample known from spectroscopy to be LINERs, and that there are no other obvious differences between these five and the rest of the LINERs, suggests that the continuum source is unobscured along our line of sight in $\sim 20\%$ of these objects. Alternatively, it can be interpreted as the fraction that is powered by photoionization, with the rest being excited by some other mechanism. We also present data for the remaining four galaxies with bright UV nuclei, one of which is a known Seyfert galaxy and three of which are H II nucleus galaxies.

In §2, we describe our observations. In §3 we present our analysis of the *HST* images, give the angular and the physical size limits as well as the UV flux, and provide the spectroscopic classification for each galaxy. Our results are discussed and summarized in §4.

## 2  Sample and Observations

The galaxies included in this paper were observed with *HST* between 1993 April and July, before the repair mission which corrected the telescope optics. The observations were done in Snapshot mode – i.e., targets were chosen from a large sample by the Space Telescope Science Institute (STScI) staff based on the convenience of their location on the sky (e.g. Bahcall et al. 1992). The brief exposure was used to fill the gaps left in the observing schedule after other science programs had been scheduled.

The sample of galaxies from which objects were chosen was defined as all galaxies in the UGC and ESO catalogs (Lauberts & Valentijn 1989) with heliocentric velocities less than 2000 km s$^{-1}$ and photographic diameters (as defined in the catalogs) greater than 6$'$. Digitized photographs of all potential target galaxies from the GASP archive at STScI were examined and the coordinates of the nucleus determined to $\sim 1''$ precision by computing the centroid of the light distribution. Ground-based CCD images were obtained for 10 galaxies in which the digitized plates were saturated in the central regions and the galactic structure was too irregular in the outer regions to determine the position of the nucleus.

For a variety of reasons, a total of 57 galaxies were removed from this initial sample of 270 galaxies during subsequent stages of the sample definition: For 8 galaxies no redshifts were available; 27 galaxies were proposed targets of other approved *HST* UV imaging programs; for 21 no nuclear position could be defined because they were too diffuse, too low in surface brightness, or were edge-on galaxies with strong dust lanes; one galaxy had a bright foreground star very near the nucleus, which would have endangered the *HST* instruments.

This selection process left a final sample of 213 galaxies from which the STScI staff chose Snapshot targets based only on scheduling convenience. An additional sample of 43 peculiar or interacting galaxies, out to a redshift of 8000 km s$^{-1}$, was also included in the sample, and will be discussed elsewhere. Two of the galaxies discussed here, NGC 1275 and NGC 5996, are from the peculiar/interacting sample. A total of 110 out of the 213 "normal" galaxies were observed while the program was active.

Data were obtained with the *HST f/96* Faint Object Camera (FOC; Paresce 1990) in its "zoomed" $1024 \times 512$-pixel mode with $0.022'' \times 0.044''$ pixels, giving a field of view of $22'' \times 22''$. The F220W filter was used. This is a broad-band filter with an effective wavelength of $\sim 2300$ Å and effective bandpass of $\sim 500$ Å. Exposures were guided with "fine-lock" or "coarse-track" guiding. The exposure time was 10 minutes per galaxy.

The images were processed by STScI's "pipeline" reduction (Baxter et al. 1994), which consists of the following steps: (1) Splitting the counts in the rectangular zoomed pixel into two square pixels ("dezooming"); (2) Correcting the geometric distortions in the image using reseau marks etched on the detector ("geometric correction"); and (3) division by a smoothed UV flat-field image to correct for detector response variations on large scales. Small-scale detector variations (e.g., scratches, reseau marks) are not corrected in the data.

## 3  Analysis

The FOC images show a variety of morphologies and UV brightnesses in the centers of the galaxies. In this paper, we focus on only nine galaxies whose images show very conspicuous bright and compact sources. In all nine cases the bright sources are within $3''$ of the center of the image. Considering the accuracy with which the coordinates of the nucleus were determined based on centroiding on the outer galactic regions in optical photographs ($\sim 1'' - 2''$), and the *HST* pointing accuracy ($\sim 1''$), the positions of all bright sources are consistent with being in the galactic nucleus. Table 1 lists the nine galaxies and their nuclear coordinates.

All our data were obtained before the *HST* repair mission at the end of 1993, and therefore are affected by spherically aberrated optics. As a result, the PSF consists of a sharp core of full width at half maximum (FWHM) $\sim 0.05''$ that contains about 15% of the light, with the rest of the light spread in a complex low-level "halo" of several arcsecond radius (Burrows et al. 1991). In the observing mode we have used, the FOC is limited in its dynamic range to 256 counts (8 bits) per zoomed pixel; additional signal causes the counts to "fold over" and start again from 0. Another problem is that the detected count rate becomes nonlinear, gradually saturating for bright sources. We estimate from our data that above a "true" rate of $\sim 0.2$ counts s$^{-1}$ pixel$^{-1}$ the detected rate is highly nonlinear and hence unreliable (see also Baxter et al. 1994). The nine sources discussed in this paper are all sufficiently bright (typically $\sim 10$ counts s$^{-1}$ pixel$^{-1}$ in the central pixels, deduced from the brightness of the halo) that the cores of their PSFs are saturated and/or nonlinear. To extract information we must rely mainly on the PSF halos. The nonlinear PSF cores can sometimes still be useful in setting upper limits on the FWHM of the source when strong limits cannot be derived from the halo structure.

Figure 1 shows sections of the *HST* images centered on the bright sources. We wish to answer the following questions: (1) Are the sources resolved or unresolved? (2) What are the UV fluxes? (3) Are there additional structures or sources near the brightest UV source? These questions are normally answered through two-dimensional PSF fitting to the images. In the present case, however, several complications exist: the central part of the PSF is nonlinear and/or folded over, the counts per pixel in the outer PSF halo are low ($\sim 1$), and small-scale variations in the detector have not been removed through flat-fielding. Furthermore, the PSF is time-variable due to "breathing" (see below). The most prominent features in the halo of the FOC PSF are the diffraction rings (Fig. 1), mainly at radii of 1.1″ and 1.8″. We find that the questions above can best be answered by analyzing the azimuthally-averaged radial profiles of the observed UV sources and comparing them to the radial profile of an empirical PSF.

To compare two images of suspected point sources, we first rebin each image by summing the counts in every 4 × 4 pixels into one new pixel, since the dezooming and the geometric correction applied to the images have made the counts in the individual pixels interdependent. We create a binned radial profile giving the mean counts per pixel as a function of radius, and the standard deviation of the mean (calculated from the distribution about the mean of the counts in the pixels in every radius bin). The amplitude of one PSF radial profile and the background beneath it are then fitted to the other PSF profile by minimizing the $\chi^2$ statistic, and the $\chi^2$ value of the best fit gives a measure of whether the two profiles are statistically consistent. The fit is done only at large radii ($r > 0.44″$) where the count rate is sufficiently low to be linear, and out to $r = 2.2″$, which includes most of the PSF flux.

The structure of the PSF is strongly dependent on wavelength (Baxter et al. 1993). It is therefore important to use model PSFs (e.g., images of stars) obtained with the F220W filter that we used for the galaxies. We examined two empirical F220W stellar PSFs observed with the FOC $f/96$. One was taken with the same observing configuration as our data, so that it too has a nonlinear and folded-over PSF core. The second was obtained with the $256 \times 256$ pixel mode that has large dynamic and linear range (but small field of view), and so is well exposed in all parts of the PSF. Comparison of the two PSFs (e.g., by "blinking" between them on a screen) shows clearly that they are different. Similarly, $\chi^2$ fitting of one radial profile to the other gives an unacceptable fit, with $\chi^2 = 16$ per degree of freedom. Comparison of the radial profiles shows that the diffraction rings are at slightly different radii in the two stellar PSFs. For the same reason, the radial profiles of both of the stellar PSFs fit poorly all nine of the observed bright UV sources. The diffraction rings in the nine objects are at larger radii than in either of the stellar PSFs. This phenomenon, of a time-variable PSF, has already been noted (Baxter et al. 1993) and is ascribed to desorption and "breathing" of the *HST* telescope structure. Desorption causes a long-term change in focus, but "breathing" apparently occurs within an orbital timescale due to temperature variations. As shown by Baxter et al. (1993), breathing involves a change in focus with a maximum amplitude of 4 $\mu$m, causing significant changes in the PSF. The stellar PSFs are therefore inadequate for fitting our data.

Since no other F220W PSFs are available, we have adopted the strategy of fitting the bright sources in our data *with each other*, with the following rationale. The exposures analyzed here were taken within several months, all with the same configuration and exposure times, so they may suffer less significantly from focus changes. By appearance, the diffraction rings of most of the UV sources seem as pronounced as those of the stellar PSFs. When any of them are convolved with a Gaussian of FWHM$\gtrsim 0.2''$, the diffraction rings in the radial profiles become very noticeably less pronounced and smoothed out. It is therefore clear that the UV sources are compact, and either unresolved or marginally resolved. Since these objects have a large range of distances, it is implausible that all of them are resolved with exactly the same FWHM. If most of the objects can be fit with each other, and there are not several objects with more pronounced diffraction rings than the others, then this is an indication that most are in fact unresolved. For those objects that are poorly fit by the other objects, we can see whether broadening of the unresolved objects serving as model PSFs can improve the fit, and thereby estimate the degree to which they are resolved. Using the amplitude of the fit to the PSF wings and the one stellar PSF that is unsaturated, we can also reconstruct the total counts in the point source.

Using this procedure, we have fit the radial profiles of each of the objects with the radial profiles of several of the other objects. The objects used for "model" PSFs are those that don't have obvious problems such as secondary sources near the bright source. We find that most of the sources are indeed well fit by the radial profiles of the other sources. By subtracting from the object being fit a properly scaled version of the object serving as the model PSF, we can then examine the residual image for additional sources and structure. Figure 2 shows the radial profile of one of the sources, NGC 4569, and the fit to it with the radial profile of another, NGC 3344.

We translate the FOC counts to a flux at 2270 Å assuming 1 count s$^{-1}$ = 1.66$\times 10^{-17}$ erg s$^{-1}$ cm$^{-2}$ Å$^{-1}$, based on the on-line calibration data available from STScI for the FOC and F220W filter, with a 25% increase in sensitivity of the 512$\times$1024 zoomed-pixel mode relative to the 512 $\times$ 512 pixel mode (Greenfield et al. 1993; Baxter et al. 1994). The uncertainty in the absolute flux scale is 10-20% (Baxter et al. 1994). We have found by integrating the F220W filter plus system transmission curve over power laws of various spectral indices that at 2270 Å the counts from a given flux are weakly dependent (differences of $\pm 3$%) on the assumed spectral shape in the UV, whether $f_\nu \propto \nu^{-1}$, as is common in weak AGNs, or $f_\nu \propto \nu^1$, which is the case for hot stars.

An additional concern in UV photometry with the FOC is the presence of "red leaks" through the F220W filter, which we define as light of wavelength $\lambda > 3200$ Å that may pass the low, but non-zero, transmission of the filter at these wavelengths. This can be a problem when observing very red sources. Recent tests indicate that the F220W transmission curve is not significantly changed from that measured before launch (Baxter et al. 1994; Hack 1994), i.e., it peaks at ~ 2200 Å and falls monotonically to the red. If a significant fraction of the F220W flux we observe in the point sources discussed in this paper were in fact caused by the red leak, we would see an unreasonably high surface brightness, of order 10 mag arcsec$^{-2}$, in the central seeing disk of ground-based optical images of these galaxies. From ground-based broad-band imaging and spectrophotometry we estimate that $f_\lambda$ at 4400 Å in the central seeing disk of these galaxies is only about 10 times greater than $f_\lambda$ at 2270 Å, while the F220W+FOC transmission decreases by a factor of about 1000 between these two wavelengths. Thus, even if all the light in the seeing disk of the ground-based data were coming from a point source, the red-leak in the *HST* data would constitute at most a few percent of the observed UV flux.

We describe below the fitting results individually for each of the bright UV sources. For each object, we also discuss the spectral classification. In most cases the optical spectra were obtained with the Hale telescope at Palomar Observatory; see Filippenko & Sargent (1985) and Ho, Filippenko, & Sargent (1994) for details. We have obtained spectra for NGC 247 and NGC 5996, which are not part of the Filippenko and Sargent sample of galaxies. Details of the spectroscopic observations, reduction, and analysis are given in Ho et al. (1994). In order to measure accurate emission-line fluxes from the spectra, which are often strongly contaminated by stellar absorption lines, we have subtracted stellar and galaxy templates using a least-squares fitting process (Rix & White 1992; Rix 1994). For several objects we have obtained spectral information from Keel (1983). A summary of the results for each galaxy appears in Table 1.

Except for two objects (NGC 247 and NGC 3344), the galaxies in which we have detected compact UV sources have been previously observed with the *International Ultraviolet Explorer (IUE)*. As a consistency check, we have compared our $f_\lambda(2270\text{Å})$ fluxes to those in the *IUE* spectra in Kinney et al. (1993) and in the IUE-ULDA compilation (Longo, Capaccioli, & Wamsteker 1992; Courvoisier, Paltani, & Wamsteker 1992). The *IUE* spectra of these relatively faint sources are particularly noisy in the 2300Å region. The *IUE* aperture is also large (equivalent to a 15.1″ diameter circular aperture), and so can include a large amount of light extraneous to the nuclear UV source. Nonetheless, we find that in all but one case the measured *HST* flux is between ~ 50 − 100% of the measured *IUE* flux. In other words, despite the large *IUE* aperture, the dominant UV flux in these objects is from the compact nucleus, consistent with what is seen in the *HST* images. The good match also confirms the reliability of our PSF measurements and the FOC photometric calibration. In the case of NGC 4736, where the *HST* image shows bright extended emission underlying the two central point sources, the *IUE* 2300Å flux is about ten times greater than that which we measure from the point sources, as expected.

Distances to the galaxies are required for making estimates of physical scales and luminosities. The distance estimates to the galaxies in our sample, aside from the usual uncertainty due to the value of the Hubble parameter $H_0$, are affected by peculiar velocities, which for such nearby galaxies can be a substantial fraction of the recession velocity. When possible, we have compiled distance estimates (generally Tully-Fisher) from the literature, or else used a distance based on a *corrected* velocity indicative of the Hubble flow at a particular location on the sky and redshift. The corrected velocity was kindly calculated for us by M. Strauss using the POTENT velocity field model (Bertschinger et al. 1990) constructed from the latest peculiar velocity data (Willick et al. 1994), and smoothed with a Gaussian window of 1200 km s$^{-1}$. In the three cases where we have used POTENT the correction to the recesion velocity is modified by up to 25%. We define $h = H_0/(100$ km s$^{-1}$ Mpc$^{-1}$).

## 3.1 Notes on Individual Objects

### 3.1.1 NGC 247 = ESO 0044−2102

This is an Sd galaxy in the Sculptor group, noted for a bright optical nucleus. The nuclear UV source in the *HST* image of this galaxy is poorly fit ($\chi^2 = 2.2$ per degree of freedom, 17 degrees of freedom) by PSFs formed from any of the other bright sources. The fit, however, degrades further if the PSFs are broadened by convolution with a Gaussian (e.g., to $\chi^2 = 3.9$ for FWHM= 0.22″). Examination of the radial profile shows that the fit is imperfect because the PSFs are different—the diffraction rings are shifted outward in the same sense as that between the stellar PSFs described above and the PSFs of the other bright sources. Based on the radial profile, the diffraction rings in this object are, however, just as sharp as in the other objects and in the stars. The image of NGC 4569, where the bright source is useful for modeling most of the other galaxies, was taken only 3 days later. This change in the PSF is evidence of a strong "breathing" effect, presumably because the thermal condition of the telescope during the NGC 247 exposure was different from that of the others. The FWHM of the saturated core is 0.13″, which constitutes an upper limit to the true FWHM. We conclude that the source is probably unresolved at FWHM$< 0.1''$, corresponding to a physical scale of $< 0.8h^{-1}$ pc at distance of $1.7h^{-1}$ Mpc (Pierce & Tully 1992).

Observations of the nucleus were made on 1993 September 11 UT with the Kast double spectrograph (Miller & Stone 1993) mounted on the Shane 3 m telescope at Lick Observatory. The data reveal a strong absorption-line spectrum indicative of a young stellar population. Subtraction of a stellar absorption line template reveals that this object has emission-line intensity ratios characteristic of an H II region, and is photoionized predominantly by OB stars; it is an "H II nucleus"

### 3.1.2 NGC 404 = UGC 718

The bright UV source in this S0 galaxy is surrounded by several point sources 3–5 mag fainter ($0.2''$ to the north; $0.8''$ PA$= -104°$; $1.1''$ PA$= 128°$, where the position-angle [PA] is measured from the north through the east) and some diffuse emission $0.2''$ to the east, with a combined flux that is non-negligible compared to the main bright source. Because of the complex structure, there is no easy way to model this object with the PSF. The saturated core of the bright source has FWHM$= 0.11''$, from which we conclude that the true FWHM is $\lesssim 0.1''$, and probably unresolved. At the distance of $1.8h^{-1}$ Mpc (Tully 1988) this corresponds to $< 0.9h^{-1}$ pc. Based on spectroscopy of the nucleus, this galaxy is a LINER with strong hydrogen Balmer absorption lines. It is possible that the low-ionization nuclear emission lines come from the bright unresolved source, while the strong young stellar features come from the surrounding structure.

### 3.1.3 NGC 1275 = UGC 2669

This is a well-studied Seyfert galaxy (e.g., Holtzman et al. 1992, and references therein) which shows evidence of a recent collision or merger. It is part of the peculiar/interacting sample we have observed. The bright UV source is well fit ($\chi^2 = 1.2$ per degree of freedom, 19 degrees of freedom) by the PSF modeled with the bright source in NGC 4569. Convolution of the PSF image with a Gaussian of FWHM$=0.22''$ degrades the fit to $\chi^2 = 2.6$, and FWHM$=0.26''$ results in $\chi^2 = 3.6$. The bright source is therefore probably unresolved with FWHM$< 0.05''$, corresponding to a physical scale of $< 13h^{-1}$ pc, assuming a distance of $52h^{-1}$ Mpc (from POTENT). There is a faint point source $0.65''$ from the bright source at PA$= -72°$, about 4.5 mag fainter than the nucleus, as well as a handful of faint sources within $6''$ south of the nucleus. The foreground F-star $5''$ NE of the nucleus noted by Hughes & Robson (1991), and which in optical exposures appears similar in brightness to the nucleus, is 3.5 mag fainter than the nucleus in the F220W ultraviolet band. We do not detect the numerous compact sources seen by Holtzman et al. (1992) in their $V$ and $R$ band $HST$ images of this galaxy, and postulated to be newly-formed globular clusters, presumably because the UV flux from these objects is faint and the sensitivity of the UV image is lower. The field of view does not include the luminous star cluster found by Shields & Filippenko (1990).

### 3.1.4 NGC 3344 = UGC 5840

This is a ringed Sbc galaxy. The nuclear UV source in the $HST$ image is well fit ($\chi^2 = 1.6$ per degree of freedom, 19 degrees of freedom) by the PSF modeled with the bright source in NGC 4569. No additional structure or sources are seen in the residual image. Convolution of the PSF image with a Gaussian of FWHM=$0.11''$ degrades the fit to $\chi^2 = 2.0$, and FWHM=$0.22''$ results in $\chi^2 = 4$. The source is therefore probably unresolved with FWHM$<0.05''$, corresponding to a physical scale of $< 1.4h^{-1}$ pc, assuming a distance of $5.6h^{-1}$ Mpc (POTENT). The nuclear spectrum is that of an H II nucleus.

### 3.1.5 NGC 4569 = M90 = UGC 7786

This is an Sab galaxy in the Virgo cluster. The bright nuclear UV source in the $HST$ image is well fit ($\chi^2 = 1.3$ per degree of freedom, 19 degrees of freedom) by the PSF modeled with the bright source in NGC 3344 or the bright source in NGC 1275. Convolution of the PSF image with a Gaussian of FWHM=$0.11''$ degrades the fit to $\chi^2 = 1.9$, and FWHM=$0.22''$ results in $\chi^2 = 4.9$. The bright source is therefore probably unresolved with FWHM$<0.05''$, corresponding to a physical scale of $< 2.2h^{-1}$ pc, assuming a distance of $9.2h^{-1}$ Mpc (Rood & Williams 1993; see, however, Stauffer, Kenney, & Young 1986, who advocate a distance of $\sim 13$ Mpc). After subtraction of the PSF, there is some residual extended emission $0.65''$ south of the bright source. Spectroscopy shows NGC 4569 to be a "transition object" between an HII nucleus and a LINER. Ho, Filippenko, & Sargent (1993) speculate that the optical spectra of such objects are powered by a composite source of ionization that is both stellar and nonstellar. The residual extended emission near the point source may originate from young stars.

### 3.1.6 NGC 4579 = M58 = UGC 7796

The nucleus of this Virgo cluster Sb galaxy has a LINER spectrum, with evidence of weak broad H$\alpha$ emission (Filippenko & Sargent 1985), as well as X-ray emission (Halpern & Steiner 1983) and a nuclear flat-spectrum radio core (Hummel et al. 1987). Based on large-aperture spectra obtained with *IUE*, Goodrich & Keel (1986) have proposed that the UV continuum is similar to the power law visible in Seyfert 2 galaxies (see, however, Kinney et al. 1993).

The radial profile of the bright source we observe in the *HST* image of this galaxy is not well fit ($\chi^2 = 2.2$, 17 degrees of freedom) by a PSF modeled with the other bright sources. The main reason for this is apparently the presence of a second, slightly extended ($\sim 0.1'' \times 0.2''$) source $0.58''$ from the bright source at PA= $72°$. The secondary source has about 15% of the flux of the bright source. We have attempted to remove this feature and then refit the bright source. The $\chi^2$ for the PSF fit to the bright source, however, does not improve, due to residual structures in the image. It is uncertain whether these structures are real or are the result of the imperfect modeling of the extended source. However, broadening of the PSF used for the fit to the bright source degrades the fit (e.g., to $\chi^2 = 3.1$ for FWHM= $0.26''$). The FWHM of the saturated core of the bright source is $0.11''$, and is an upper limit to the true FWHM of the source. We therefore estimate that the bright source is unresolved, with FWHM$< 0.1''$, which is $< 6.8 h^{-1}$ pc at a distance of $13.9 h^{-1}$ Mpc (Pierce & Tully 1988).

### 3.1.7 NGC 4736 = M 94= UGC 7996

This Sab galaxy has been noted elsewhere for a ring of H II regions at a radius of $50''$ ($\sim 0.7$ Kpc), with strong star formation, red arcs at $\sim 15''$ ($\sim 0.2$ Kpc) radius, a high surface-brightness nuclear region, high far-infrared bulge emission, a LINER nucleus, and an extended ($5''$; $\sim 100$ pc) radio source centered on the nucleus (Kinney et al. 1993; Smith et al. 1994, and references therein).

The *HST* image of the nucleus shows *two* compact sources of approximately equal brightness separated by $2.5''$ with north-south orientation (PA= $-3°$; see Fig.1). There is high surface brightness diffuse emission centered on the southern source, presumably corresponding to the optical nucleus. The two compact sources are bright enough ($\sim 0.3$ observed counts s$^{-1}$ in the peaks) that nonlinearity affects their PSF cores, but faint enough, and the background from the diffuse emission high enough, that the PSF diffraction rings cannot be seen in the radial profile. The system is therefore difficult to model by PSF fitting. The FWHM of the nonlinear PSF cores of the two bright sources are $\sim 0.1''$, and so the sources are probably both unresolved at a physical scale of $< 1.6h^{-1}$ pc, assuming a distance of $3.2h^{-1}$ Mpc (Tully 1988). Also remarkable in this image are three large concentric arcs at radii of approximately $2''$, $4''$, and $6''$ south and west from the bright central emission, suggestive of bow shocks or tidal arms. The system appears to be in the final stages of a merger. The Filippenko & Sargent (1985) spectrum of the nucleus shows a transition type object, with features of LINER and H II spectra and strong Balmer-line absorption. Following starlight subtraction we find line intensity ratios similar to those measured by Keel (1983), confirming that the source at the position of the optical nucleus is a LINER.

### 3.1.8 NGC 5055 = M63 = UGC 8334

The bright UV source in the nucleus of this Sb galaxy is apparently resolved by *HST*; its image is not well fit by the PSF modeled using the other bright sources, due to an excess of flux in the inner part. Limiting the fit to radii of $0.62''$ to $2.2''$ in order to avoid the excess flux, we obtain a reduced $\chi^2 = 2.5$ with a PSF based on the other bright sources. The fit improves to an acceptable $\chi^2 = 1.5$ (16 degrees of freedom) if the PSF is convolved with a Gaussian of FWHM$= 0.22''$, and degrades back to $\chi^2 = 2.4$ for FWHM$= 0.33''$. The FWHM of the saturated core is about $0.2''$, and so does not give a different limit. We conclude that the bright source in this galaxy is resolved, with FWHM$\approx 0.2''$, corresponding to a physical scale of $6.0h^{-1}$ pc assuming a distance of $6.2h^{-1}$ Mpc (Phillips 1993). Optical spectra show this nucleus to have weak LINER characteristics with strong H$\alpha$ absorption, suggestive of a hot stellar population. We may be seeing a resolved star cluster, on which is superposed a possibly unresolved LINER nucleus. No additional sources are seen around the bright nucleus. This is the only bright UV source in our sample that is clearly not point-like at the *HST* resolution (apart from additional surrounding faint structures and point sources, which appear in several of the galaxies).

### 3.1.9 NGC 5996 = UGC 10033

This is a disturbed-looking galaxy at a redshift of 0.01, and is part of the peculiar/interacting sample we have observed. Optical and UV spectroscopy (Balzano 1983; Kinney et al. 1993) indicate starburst activity. The bright UV source in our *HST* image is fairly well fit ($\chi^2 = 1.7$ per degree of freedom, 20 degrees of freedom) by the PSF modeled with the bright source in NGC 4569. Convolution of the PSF image with a Gaussian of FWHM=0.22″ degrades the fit to $\chi^2 = 2.8$, and FWHM=0.26″ results in $\chi^2 = 3.7$. The bright source is therefore probably unresolved with FWHM< 0.05″, corresponding to a physical scale of $< 7h^{-1}$ pc, assuming a distance of $27h^{-1}$ Mpc (POTENT). Three additional sources appear within 2″ of the bright nucleus: at 1.4″ PA= $-106°$, at 1.6″ PA= $-117°$, and at 0.55″ PA= $55°$. There might be one more source about 0.13″ east of the nucleus, but it is superposed on the nonlinear core of the bright nucleus. The additional sources are sufficiently bright that their cores are nonlinear, yet too faint to permit a measurement their halo structure on the background of the bright PSF halo. The FWHMs of the three sources that are not too close to the nucleus suggest that each one is unresolved. Each is about 4 magnitudes fainter than the bright nucleus. Our optical (3100 − 8000 Å) spectra confirm that this is a classic starburst nucleus.

## 4  Discussion and Conclusions

The UV sources analyzed in this paper are much fainter than "classical" AGNs. The fluxes given in Table 1 correspond to AB 2270 Å magnitudes of 16 to 20. If the LINERs among these objects have $f_\nu \propto \nu^{-1}$ power-law continua, their $B$ magnitudes will be 16 − 20 which, at the distances of the galaxies, gives absolute magnitudes (see Table 1) in the range $M_B = -8$ to $-14.5$ (i.e., $2 \times 10^5 - 10^8 h^{-2} L_{B\odot}$). If these nuclei are AGNs, then they are among the least luminous known (see §1).

Ho et al. (1993) have analyzed and modeled the spectra of 13 LINERs, and concluded that photoionization by an AGN-like power-law continuum can best reproduce the observed optical emission-line properties. Our detection of UV continuum emission (no strong lines are expected in the wavelength interval we have observed) supports the hypothesis that photoionization is the excitation mechanism of at least some LINERs. The unresolved nature of the bright UV sources we detect, and the small physical dimensions which are implied, indicate that the sources are either AGNs (i.e., nonstellar in nature, e.g., powered by accretion onto a collapsed object), or dense star clusters. The latter possibility cannot be excluded based on the present data; indeed, two of the sources that are clearly unresolved (NGC 3344 and NGC 5996) are not LINERs but starburst nuclei. The Milky Way also has a young compact nuclear star cluster, of radius less than 1 pc (e.g. Eckart et al. 1994). The large-aperture, low signal-to-noise *IUE* spectra of these objects sometimes show UV-absorption features. The spectra are, however, too noisy and have and too-low resolution to tell whether these are signatures of a stellar cluster, or gas-phase absorption in the ISM of the Milky Way and the host galaxy. Spectroscopy of the UV sources with *HST* can provide the answers. We find that most of the LINERs described here have, in addition to the bright UV source, diffuse and point-like UV sources in the immediate vicinity of the nucleus. These sources could cause the stellar signatures that are often mixed into optical LINER spectra, both as early-type stellar absorption features and as some emission-line intensity ratios characteristic of gas photoionized by hot stars.

To test the hypothesis that the UV-bright LINERs are photoionized, we have compared the ionizing photon flux to the H$\alpha$ photon flux in the five LINERs, assuming the UV continuum can be extrapolated from 2270 Å, where we measure it, to beyond the Lyman limit as a $\nu^{-1}$ power law (as is commonly observed in low-luminosity AGNs). The ratio of Lyman continuum to H$\alpha$ photons is then

$$N_{ion}/N_{H\alpha} = \frac{f_\lambda(2270\text{Å}) \times 912\text{Å}}{f(\text{H}\alpha)} \times \frac{2270\text{Å}}{6563\text{Å}}.$$

This estimate is a lower limit to the true ratio, because correction for any Galactic or internal reddening will enhance the UV flux relative to the H$\alpha$ flux. We have taken the H$\alpha$ fluxes for four of the LINERs from Keel (1983), who obtained spectra through wide apertures (4.7″ and 8.1″) that include most of the nuclear flux. For NGC 5055, which Keel (1983) did not observe, we have used the Filippenko & Sargent (1985) narrow-slit H$\alpha$ flux, scaled up to correct for the small aperture by comparing the Keel and the Filippenko & Sargent fluxes for NGC 4736. (NGC 4736 was observed under the same conditions as NGC 5055.) Table 1 lists the $N_{ion}/N_{H\alpha}$ photon ratio for the five LINERs.

Assuming, for simplicity, Case B recombination conditions, the ratio of the total recombination coefficient to the H$\alpha$ recombination coefficient is 2.2 (Osterbrock 1989); in other words, 0.45 of all hydrogen recombinations lead to the emission of an H$\alpha$ photon. From Table 1, $N_{ion}/N_{H\alpha} > 2.2$ for the four LINERs for which we can measure the UV flux. The strength of the emission lines can therefore be accounted for with photoionization by an extrapolation of the observed UV continuum. It is remarkable that $N_{ion}/N_{H\alpha}$ is between 3–7, i.e., constant to within a factor of 2, while the UV and H$\alpha$ luminosities among the objects span about three orders of magnitude. (In other words, there is an excellent correlation between the observed UV flux and the H$\alpha$ flux). Note also that, after accounting for the emission-line strength, there is enough $\sim 2300$Å flux to explain *some* of the UV continuum as arising from stars, in a combined AGN + nuclear-star-cluster scenario. Again, *HST* spectra can discern the signatures of such a cluster.

Ho et al. (1993) also examined shock heating models as an alternative for explaining LINERs. Although shock models did not give good fits to the data, Ho et al. pointed out that this may be a shortcoming of the limited range of parameters that was calculated. One fairly secure prediction of shock heating models is, however, that little UV radiation should be produced, since the excitation is mechanical. Our detection of bright UV sources in five LINERs therefore constitutes strong evidence against the shock scenario, at least in these five objects.

Most of the Northern-hemisphere galaxies in our sample are common to the Filippenko & Sargent (1985) sample of galaxies. For 26 of the common galaxies, spectral analysis by Ho et al. (1994) shows LINER, or LINER + H II mixture properties. The five LINERs discussed here are the only ones among these 26 showing compact central UV sources; the remaining galaxies display morphologies similar to the other, non-LINER, galaxies in our sample. One interpretation for this is that an ionizing UV continuum source exists in all LINERs but is unobscured by dust along our line of sight in $\sim 5/26$, or 20%, of LINERs. Alternatively, it may be that LINERs are not powered by just one type of excitation mechanism (photoionization), and that some other mechanism (e.g. shocks) produces the lines in 80% of the cases.

To test the obscuration hypothesis, we have measured the strongest optical emission lines in the spectra of the LINERs common to the two samples. We plot in Figure 3 the diagnostic line-ratio diagrams (following Veilleux & Osterbrock 1987), for all the LINERs, from which we can examine whether there are spectral differences between the UV-bright and UV-dark LINERs. Because of the strong starlight contamination in the spectra of many of these objects, and the resulting absorption-line residuals left after the stellar template subtraction, H$\beta$, [OIII]$\lambda$5007, and [OI]$\lambda$6300 are sometimes undetected. In such cases, where possible we have taken the line fluxes from Keel (1983), or assumed H$\beta$ =H$\alpha$/3, or excluded the object from the diagnostic diagram.

Figure 3 shows that the objects as a whole have line-intensity ratios typical of LINERs (see, e.g., Ho et al. 1993). There is no clear segregation between the UV-bright LINERs (filled circles) and the UV-dark ones (empty circles). The UV-bright LINERs have a somewhat lower [OIII]$\lambda$5007/ H$\beta$ ratio, but this may be the result of the fact that H$\beta$ is undetected in three of the five, and we have assumed H$\beta$=H$\alpha$/3. If the spectra are reddened (as is often the case), the H$\beta$ =H$\alpha$/3 assumption will artificially lower the [OIII]$\lambda$5007/ H$\beta$ ratio. We conclude that there is no obvious distinction in the optical spectral properties of the UV-bright and UV-dark LINERs, consistent with the hypothesis that they are one type of object. This test can be extended to weaker lines with higher S/N spectra, as well as to other spectral regions.

To summarize our main results:
1. We have detected a bright compact nuclear UV source in five LINER galaxies.
2. The brightness of the UV source, when extrapolated to ionizing energies, is sufficient to explain the strength of the hydrogen lines through photoionization.
3. Only about 20% of all LINERs in a complete sample of galaxies display a nuclear UV source. This may mean that all LINERs are excited by photoionization (stellar or AGN) with the UV continuum source being obscured along our line of sight in 80% of the cases. This hypothesis is supported by the lack of obvious spectral differences between the UV-bright and UV-dark LINERs. Alternatively, photoionization may be responsible for the emission lines in only 20% of the cases, with other mechanisms, such as shocks, operating in the rest, but producing similar optical spectra.
4. The UV continuum sources are unresolved, $\lesssim 0.1''$, in all but one case. This limits the AGN continuum-emitting region or dense young nuclear star cluster to a maximum physical scale of several parsecs.
5. The bright UV point sources are generally surrounded by faint diffuse emission and point sources. These may be the origin of the young stellar features often superposed on LINER spectra.

The bright UV sources we have detected here can serve as excellent targets for future UV spectroscopic studies of LINERs. UV spectroscopy with *HST* can reveal the continuum shape, probe the multiple UV components when present, and test for the existence of additional spectral features that can discriminate between the various proposed driving mechanisms of LINERs.

**Acknowledgements** We thank the following individuals for help, comments, and advice: A.J. Barth, D. Baxter, P. Greenfield, W. Hack, T. Matheson, A. Sternberg, M. Strauss, B. Wills, and the referee, W. Keel. This research has made use of the NASA/IPAC Extragalactic Database (NED), which is operated by JPL, Caltech, under contract with NASA. This work was supported by grant GO-3519 (D.M., A.V.F.) and Hubble Fellowship HF-1025 (H.-W.R.) from the Space Telescope Science Institute which is operated by AURA under NASA contract NAS 5-26555, by NASA grant NAG5-1618 (J.N.B., D.P.S), and by NSF grant AST-8957063 (A.V.F., L.C.H.).

Table 1: Bright UV Sources

| Galaxy R.A.(J2000) Dec.(J2000) | $f_\lambda(2270\text{Å})^*$ $N_{ion}/N_{H\alpha}$ | FWHM | Size (pc) Dist.(Mpc) | Spectral Classif. $M_B^\dagger$ | Comments |
|---|---|---|---|---|---|
| **NGC 247** 00:47:08.2 $-20°45'37''$ | $6.2 \times 10^{-16}$ | $< 0.13''$ | $< 0.8h^{-1}$ $1.7h^{-1}$ | H II $-6.5$ | |
| **NGC 404** 01:09:26.9 $+35°43'04''$ | $1.8 \times 10^{-15}$ 5.2 | $< 0.11''$ | $< 1h^{-1}$ $1.8h^{-1}$ | LINER $-9.2$ | Several surrounding point sources and diffuse emission. |
| **NGC 1275** 03:19:48.0 $+41°30'44''$ | $1.8 \times 10^{-15}$ | $< 0.05''$ | $< 13h^{-1}$ $52h^{-1}$ | Seyfert $-16.5$ | Additional faint point source. |
| **NGC 3344** 10:43:31.0 $+24°55'20''$ | $4.2 \times 10^{-15}$ | $< 0.05''$ | $< 1.5h^{-1}$ $5.6h^{-1}$ | H II $-11.2$ | |
| **NGC 4569** 12:36:49.9 $+13°09'44''$ | $1.0 \times 10^{-14}$ 6.9 | $< 0.05''$ | $< 2.2h^{-1}$ $9.2h^{-1}$ | LINER $-14.6$ | Some faint extended emission. |
| **NGC 4579** 12:37:43.5 $+11°49'06''$ | $1.1 \times 10^{-15}$ 2.8 | $< 0.11''$ | $< 7h^{-1}$ $14h^{-1}$ | LINER $-13.1$ | Additional extended source, 2 mags fainter. |
| **NGC 4736** 12:50:53.0 $+41°07'12''$ | $> 2.0 \times 10^{-16}$ $> 1$ | $< 0.11''$ | $< 1.6h^{-1}$ $3.2h^{-1}$ | LINER $< -8.1$ | Two equally bright point sources; $2'' - 6''$ radius arcs. |
| **NGC 5055** 13:15:49.2 $+42°01'50''$ | $1.0 \times 10^{-15}$ 5.2 | $0.2''$ | $6h^{-1}$ $6.2h^{-1}$ | LINER $-11.2$ | Resolved. |
| **NGC 5996** 15:46:58.8 $+17°53'03''$ | $6.3 \times 10^{-15}$ | $< 0.05''$ | $< 7h^{-1}$ $27h^{-1}$ | H II $-15.0$ | 3-4 additional faint point sources. |

$^*$ In erg s$^{-1}$ cm$^{-2}$ Å$^{-1}$.

$^\dagger$ Calculated assuming $f_\nu \propto \nu^{-1}$ for the LINERs and for the Seyfert nucleus, $f_\nu \propto \nu^1$ for the H II nuclei, no reddening, and $h = 1$.

# References

Bahcall, J. N., Maoz, D., Doxsey, R., Schneider, D. P., Bahcall, N. A., Lahav, O. & Yanny, B. 1992, ApJ, 387, 56

Balzano, V. A. 1983, ApJ, 268, 602

Baxter, D. A., Greenfield, P. E., Hack, W., Nota, A. Jedrzejewski, R. I., and Paresce, F. 1993, STScI preprint No. 751

Baxter, D. A., Gilmore, D., Greenfield, P. E., Hack, W., Hodge, P., Jedrzejewski, R. I., & Nota, A. 1994, in "HST Data Handbook", ed. S. Baum (Baltimore: STScI)

Bertschinger, E. et al. 1990, ApJ, 364, 370

Bohlin, R. C., Cornett, R. H., Hill, J. K, Hill, R. S., O'Connell, R. W., and Stecher, T. P. 1985, ApJ, 298, L37

Burrows, C. J., et al. 1991, ApJ, 369, L21

Coleman, G. D., Wu C. C., & Weedman, D. W. 1980, ApJS, 43, 393

Courvoisier, T. J. -L., Paltani, S. & Wamsteker, W. 1992, "IUE-ULDA Access Guide No. 4, Active Galactic Nuclei" (Noordwijk: ESA)

Eckart, A., Genzel, R., Hoffman, R., Sams, B., & Tacconi-Garman, L. 1993, ApJ, 407, 77

Filippenko, A. V. 1989, in "Active Galactic Nuclei", eds. D. E. Osterbrock & J. S. Miller (Dordrecht: Kluwer), 495

Filippenko, A. V., Ho, L. C., & Sargent W. L. W. 1993, ApJ, 410, L75

Filippenko, A. V., & Sargent, W. L. W 1985, ApJS, 57, 503

Filippenko, A. V., & Sargent, W. L. W. 1989, ApJ, 342, L11

Filippenko, A. V., & Terlevich, R. 1992, ApJ, 397, L79

Fosbury, R. A. E., Mebold, U., Goss, W. M., & Dopita, M. A. 1978, MNRAS, 183, 549

Goodrich, R. W., & Keel, W. C. 1986, ApJ, 305, 148

Greenfield, P., et al. 1993, STScI preprint No. 751

Hack, W. 1994, private communication

Halpern, J. P., & Steiner, J. E. 1983, ApJ, 269, 37

Heckman, T. 1980, A&A, 87, 152

Ho, L. C., Filippenko, A. V., & Sargent, W. L. W. 1993, ApJ, 417, 63

Ho, L. C., Filippenko, A. V., & Sargent, W. L. W. 1994, in preparation

Holtzman, J. A., et al. 1992, AJ, 103, 691

Hughes, D. H., & Robson, E. I. 1991, MNRAS, 249, 560

Hummel,E., Van Der Hulst, J. M., Keel W. C., & Kennicutt JR. R. C. 1987, A&AS, 70, 517

Keel, W. C. 1983, ApJ, 269, 466

Kinney, A. L., Bohlin, R. C., Calzetti, D., Panagia, N., & Wyse, R. F. G. 1993, ApJS, 86, 5

Koski, A. T., & Osterbrock, D. E. 1976, ApJ 203, L49

Lauberts, A. & Valentijn, E. A. 1989, "The Surface Photometry Catalogue of the ESO-Uppsala Galaxies" (Garching: ESO)

Longo, G., Capaccioli, M., & Wamsteker, W. 1992, "IUE-ULDA Access Guide No. 3, Normal Galaxies" (Noordwijk: ESA)

Miller, J. S. & Stone, R. P. S. 1993, Lick Obs. Tech. Rep., No. 66

Oke, J. B. & Gunn, J. E. 1983, ApJ, 266, 713

Osterbrock, D. E. 1989, "Astrophysics of Gaseous Nebulae and Active Galactic Nuclei" (Mill Valley, California: University Science Books)

Osterbrock, D. E., & Martel, A. 1993, ApJ, 414, 552

Paresce, F. 1990, "Faint Object Camera Instrument Handbook" (Baltimore: STScI)

Phillips, M. M. 1993, ApJ, 413, L105

Pierce, M. J. & Tully, R. B. 1988, ApJ, 330, 579

Pierce, M. J. & Tully, R. B. 1992, ApJ, 387, 47

Rix, H.-W., & White, S. D. M. 1992, MNRAS, 254, 389

Rix, H.-W., Kennicut, R., Braun, R., & Walterbos, R. 1994, ApJ, submitted

Rood, H. J. & Williams, B. A. 1993, MNRAS, 263, 211

Shields, J. C. 1992, ApJ, 399, L27

Shields, J. C., & Filippenko, A. V. 1990, ApJ, 353, L7

Smith, B. J., Harvey, P. M., Colome, C., Zhang, C. Y., DiFrancesco, J., & Pogge, R. W. 1994, ApJ, 425, 91

Stauffer, J. R. 1982, ApJ, 262, 66

Stauffer, J. R., Kenney, J. D., & Young, J. S. 1986, AJ, 91, 1286

Terlevich, R. & Melnick, J. 1985, MNRAS, 213, 841

Tully, R. B. 1988, "Nearby Galaxies Catalog" (Cambridge: Cambridge University Press)

Veilleux, S., & Osterbrock, D. E. 1987, ApJS, 63, 295

Willick, J., et al. 1994, in preparation

# Figure Captions

**Figure 1.** Sections of the *HST* images centered on the bright UV source of each galaxy. Each section is 6.6″ on a side. The orientation is marked on each section. The same gray scale is used for all of the sections. As a result, the diffraction rings and other PSF halo structure are conspicuous among the brighter objects, while the sharp PSF core is visible among the fainter objects. The central part of the NGC 4569 image also shows the effect of counts "folding over".

**Figure 2.** Example of the radial profile fitting used to test whether the bright point sources are resolved. The dots with error bars having broad tick-marks at their ends show the radial light distribution of the bright central UV source in NGC 4569 (mean counts per pixel and the uncertainty of the mean, as a function of radius in arcseconds). The error bars without dots are the radial light distribution of the central source in NGC 3344, after it has been scaled to minimize the $\chi^2$ between the two distributions. After scaling by a factor of 2.26 and subtraction of a constant of 0.35 counts pixel$^{-1}$ (which accounts for differences in background between the two images), the source in NGC 3344 fits the source in NGC 4569 with an acceptable $\chi^2 = 1.3$ per degree of freedom for 19 degrees of freedom (21 bins minus two free parameters). The acceptable fit indicates that the central UV sources in both galaxies are resolved to the same degree, and suggests that both are probably unresolved. The local maxima in the radial profiles at 1.1″ and at 1.8″ are at the radii of the bright diffraction rings (c.f. Fig. 1). The maxima become less pronounced as a source becomes more resolved.

**Figure 3.** Diagnostic line-ratio diagrams for LINERs common to the *HST* sample and the Filippenko & Sargent (1985) sample. Filled circles denote the five LINERs that contain detected nuclear UV sources, empty circles show the other, UV-dark LINERs. No obvious segregation exists between the UV-bright and UV-dark LINERs, suggesting that they may be one class of objects, with the UV continuum usually obscured. The right-hand panel does not include the UV-bright LINER NGC 5055, and several other objects for which we have only upper limits on the [OI] $\lambda$6300 flux.

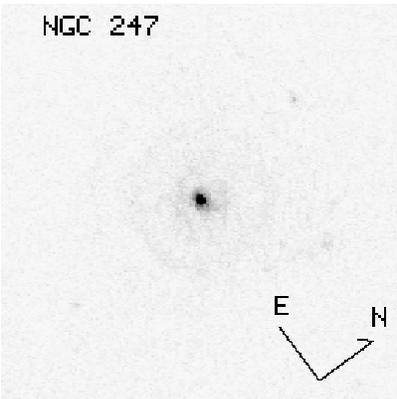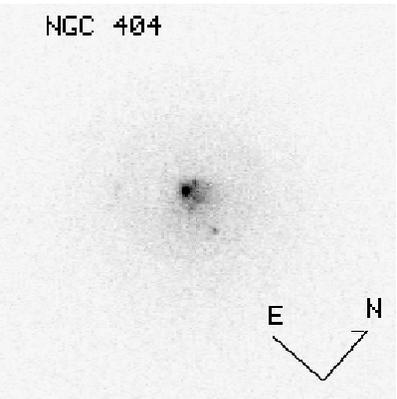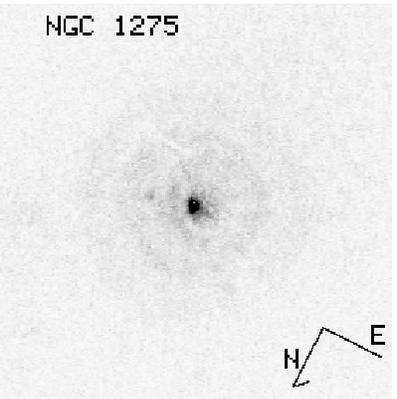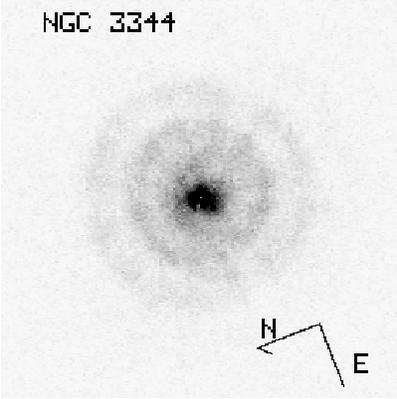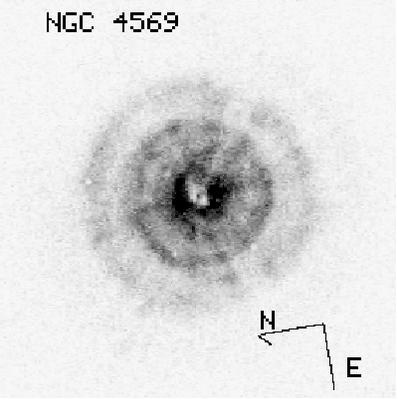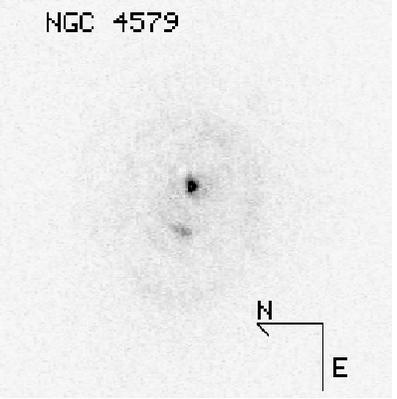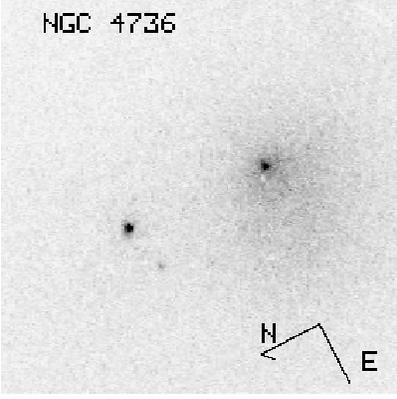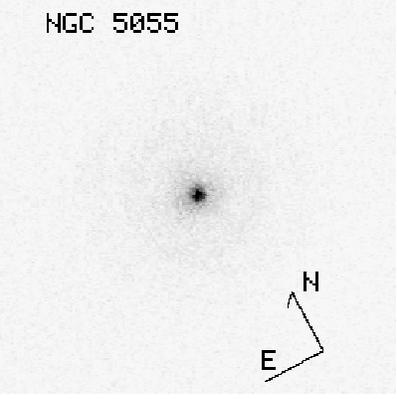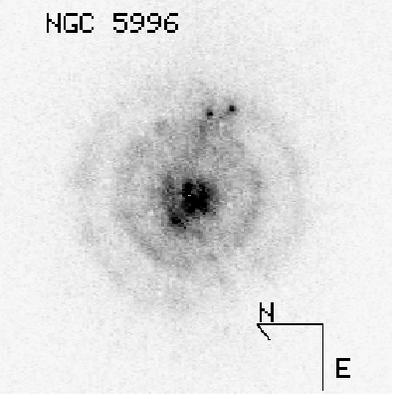

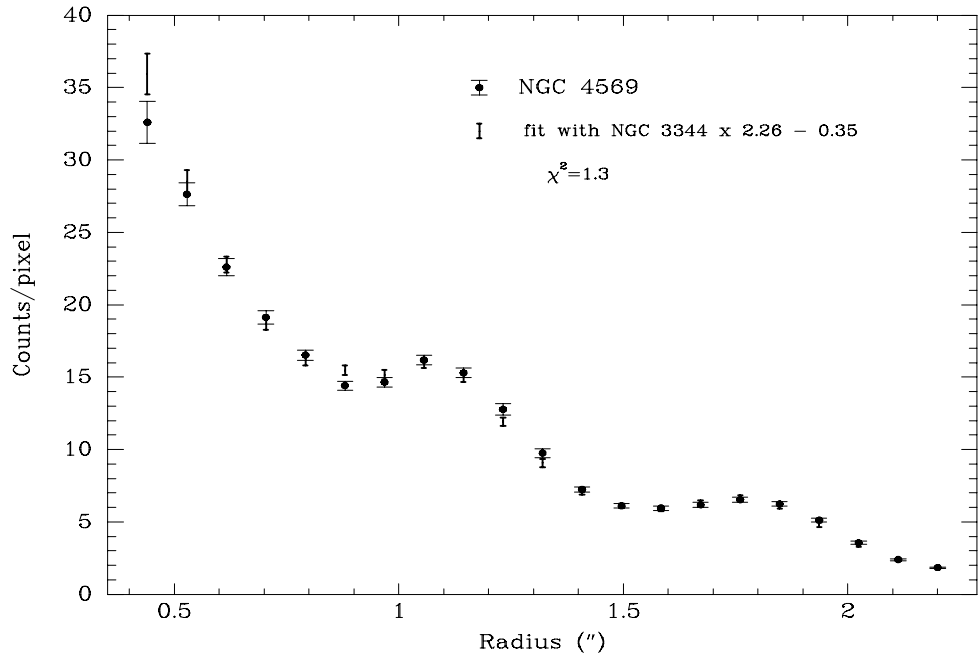

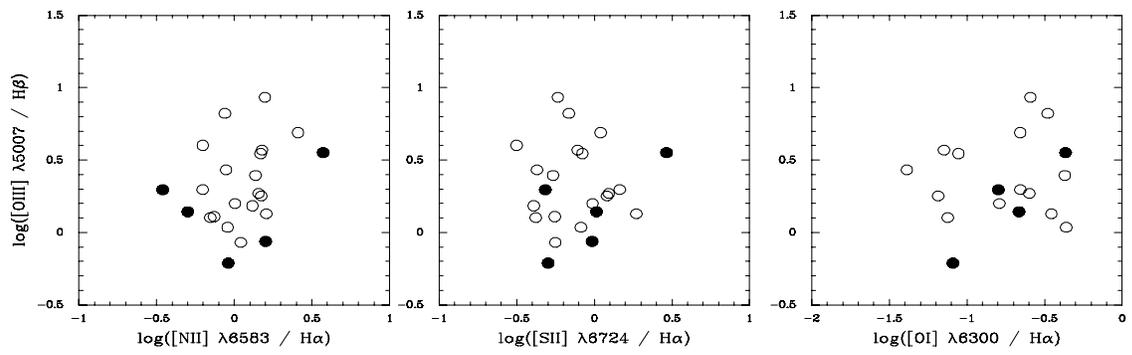